\documentclass{PoS}

\title{Testing the $\mu-e$ universality with $K^{\pm}\to l^{\pm}\nu $ decays}

\ShortTitle{Testing the $\mu-e$ universality with $K^{\pm}\to l^{\pm}\nu $ decays }

\author{\speaker{Venelin Kozhuharov}\thanks{On behalf of the NA48/2 collaboration.}\\
        University of Sofia "St. Kl. Ohridski"\\
        E-mail: \email{Venelin.Kozhuharov@cern.ch}}

\abstract{   The ratio $R_K = \Gamma(K^{\pm}\to e^{\pm}\nu) /
 \Gamma(K^{\pm}\to \mu^{\pm}\nu) $ provides a very powerful probe for
 the weak interactions structure.
 This ratio of decay rates is calculated with very high precision within 
the Standard Model but the
 corrections due to the presence of New Physics could be as high as 3\%.
 The data obtained by the NA48 experiment at the CERN SPS accelerator during a 56 hours special run in 2004
 has been  analyzed.
The precision of the preliminary result for $R_K$ is two times better than the world average
 but is still insufficient to probe the existence of physics Beyond the 
Standard Model.
 Currently the experiment is taking data dedicated to the 
sub-percent precision measurement of $R_K$. }

\FullConference{KAON International Conference\\
		 May 21-25 2007\\
		 Laboratori Nazionali di Frascati dell'INFN, Rome, Italy}

\begin{document}

\newcommand{\ke}{$K_{e2}~$}
\newcommand{\km}{$K_{\mu 2}~$}
\newcommand{\Ke}{K_{e2}~}
\newcommand{\Km}{K_{\mu 2}~}

\section{Introduction}
In the Standard Model (SM) the weak interactions of the leptons are described by a unique coupling constant $G_F$. 
This universality can be tested by investigating the leptonic decays of light mesons like
 $K^{\pm}\to l^{\pm}\nu $ decays. They proceed as tree level processes within the SM and the ratio between of the decay widths of the electronic and muonic mode $R_K= \Gamma(Ke2) / \Gamma(K\mu 2)$ can be written as \cite{Finkemeier}
\begin{equation}
R_K=\frac{m_e^2}{m_{\mu}^2} \left(  \frac{m_K^2 - m_e^2}{m_K^2 - m_{\mu}^2} \right)   (1+\delta R_K)
\end{equation}
where the term $ \delta R_K = -(3.78 \pm 0.04) \% $ represents the radiative corrections. In the ratio $R_K$ the theoretical uncertainties on the hadronic matrix element cancel resulting in a precise prediction $R_K = (2.472 \pm 0.001) \times 10^{-5} $.

Considering different extensions of the SM (like models with Lepton Flavour Violation, different SUSY scenarios) a constructive or destructive contribution to $R_K$ as high as 3\% could be achieved \cite{Masiero}. 

The value of $R_K$ has been measured by three experiments in the 70s \cite{ke2prev} but the precision of the combined result ($ R_K = (2.45 \pm 0.11 ) \times 10^{-5} $) doesn't allow to perform a significant test of the lepton universality within the SM.

\section{The NA48/2 experiment}
The NA48/2 experiment was primarily designed to measure the charge asymmetry in the $K^{\pm}\to3\pi$ decays in order to probe for CP violation in the charged kaon system \cite{na482}.
%, \cite{kaon07eg}. 
Data were taken in 2003 and 2004, providing approximately $4 \times 10^9$   $K^{\pm}\to \pi^{\pm}\pi^+\pi^- $ decays. Apart from the nominal running conditions the data taking also included  two special runs with reduced beam intensity and special trigger setup devoted to the investigation of kaon semileptonic decays. 

The huge statistics allows to study rare kaon decays with high precision. For the measurement of $R_K$ described below only the data taken during 56 hours special run in 2004  was used. 

Kaons at the NA48/2 experiment are produced by a $400$ GeV/c primary proton beam from the CERN SPS hitting a beryllium target. A system of collimators and achromats selects particles with mean momentum of $60$ GeV/c
 with both positive and negative charge. The momentum of the particles is measured by a beam spectrometer. The simultaneous $K^{\pm}$ beams enter a 114 m long evacuated decay region closed by a Kevlar window. The decays products are detected by the NA48 detector complex including:

{\bf - Magnetic spectrometer (DCH):}
		Composed by four drift chambers and a dipole magnet deflecting the charged particles in the horizontal plane. The resulting resolution on the momentum measurement is $\delta p/p = 1\% \oplus 0.044\% p$ [GeV/c].

{\bf - Charged hodoscope (HOD):}
	Fast scintillator detector with time resolution of $150$ ps used in the trigger system.

{\bf - Electromagnetic Calorimeter (LKR):}
	A Liquid Krypton calorimeter of 27 radiation lengths used to measure the energy of the electrons and photons with resolution of $ \delta E/E = 3.2\%/ \sqrt{E} \oplus 9\%/ E \oplus  0.42\%$ [GeV].

The mean beam axis is close to the $Z$ axis of the experiment. A more detailed description of the experimental setup can be found elsewhere \cite{NA48setup}

\section{Data analysis}

\subsection{Event selection}

Due to the reduced beam intensity only highly efficient minimum bias triggers were used to collect the data. 
\ke events were triggered by a time coincidence of 
signal from the HOD compatible with one charged track and energy deposition in the LKR higher than $10$ GeV. 
Since \km is dominant decay mode of the charged kaons they were collected using only the 
HOD signal downscaled by a factor of 50. 

In the final state $Kl2$ events contain only one observable particles, the charged lepton, 
and this similarity allowed to minimise the differences in the offline selection criteria. 

{\bf Geometry and kinematics:}
The charged particle was required to pass through the DCH and HOD and LKR and the event had to contain no extra tracks or clusters in the LKR with energy higher than $2$ GeV. 
Based on the closest distance approach between the extrapolated track from the DCH and the beam axis, 
the Z coordinate of the decay vertex was computed and was required to be between $2000$ cm and $ 7000$ cm (the decay region starts at $-1800$ cm). 
The track momentum had to be $15$ GeV/c $< P_{Trk} < 50$ GeV/c. The interval was split into 7 bins of $5$ GeV/c and the analysis was performed independently in each of the bins.

{\bf Particle identification (PID):}
For the separation between electrons and muons the ratio $E/pc$ was used, where $p$ is the track momentum measured from the DCH and $E$ is the reconstructed energy of the particle in the LKR. For electrons this variable peaks at 1 and the condition of $E/pc>0.95$ was used. Muons were identified as particles with $E/pc<0.2$. 

For a given assumption on the lepton mass, the missing mass squared $M_{miss}^2 = (P_K - P_l)^2$ of the event was computed using $60$ GeV/c along the beam axis for the kaon momentum. 
% as $M_{miss}^2 = (P_K - P_l)^2$, where 
In the case of $Kl2$ event it had to be consistent with the mass of the neutrino. 

\begin{figure}[!htb]
  \begin{minipage}[t]{0.47\textwidth}
    \center\includegraphics[width=70mm]{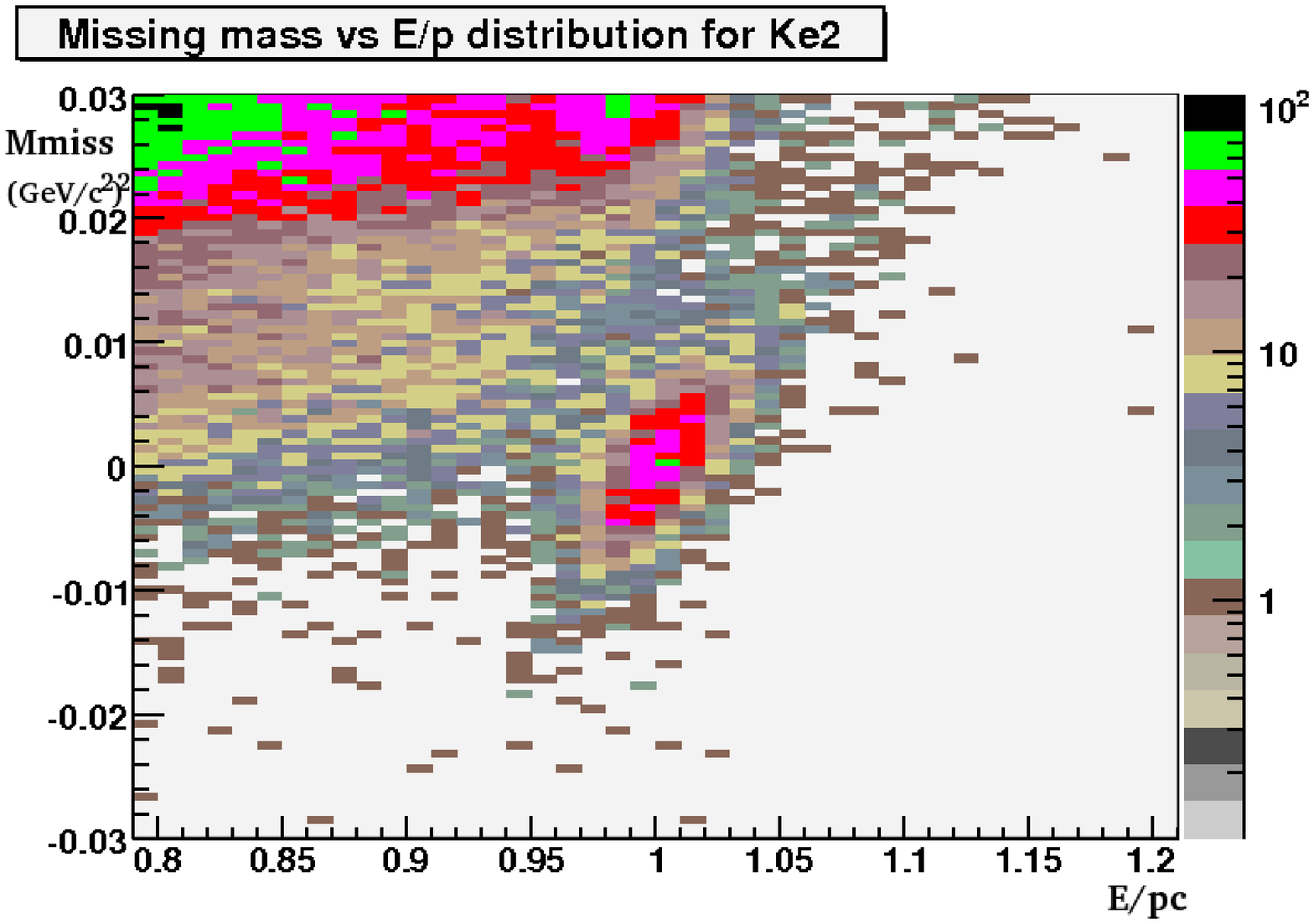}
    \caption{Missing mass versus the $E/pc$ distribution for the $Ke2$ candidates before $M_{miss}$ and $E/pc$ cuts. A logarithmic scale is used for the event density in order to see the background contribution }
    \label{ke2mmiss}
  \end{minipage}\hfill
  \begin{minipage}[t]{0.47\textwidth}
    \center\includegraphics[width=70mm]{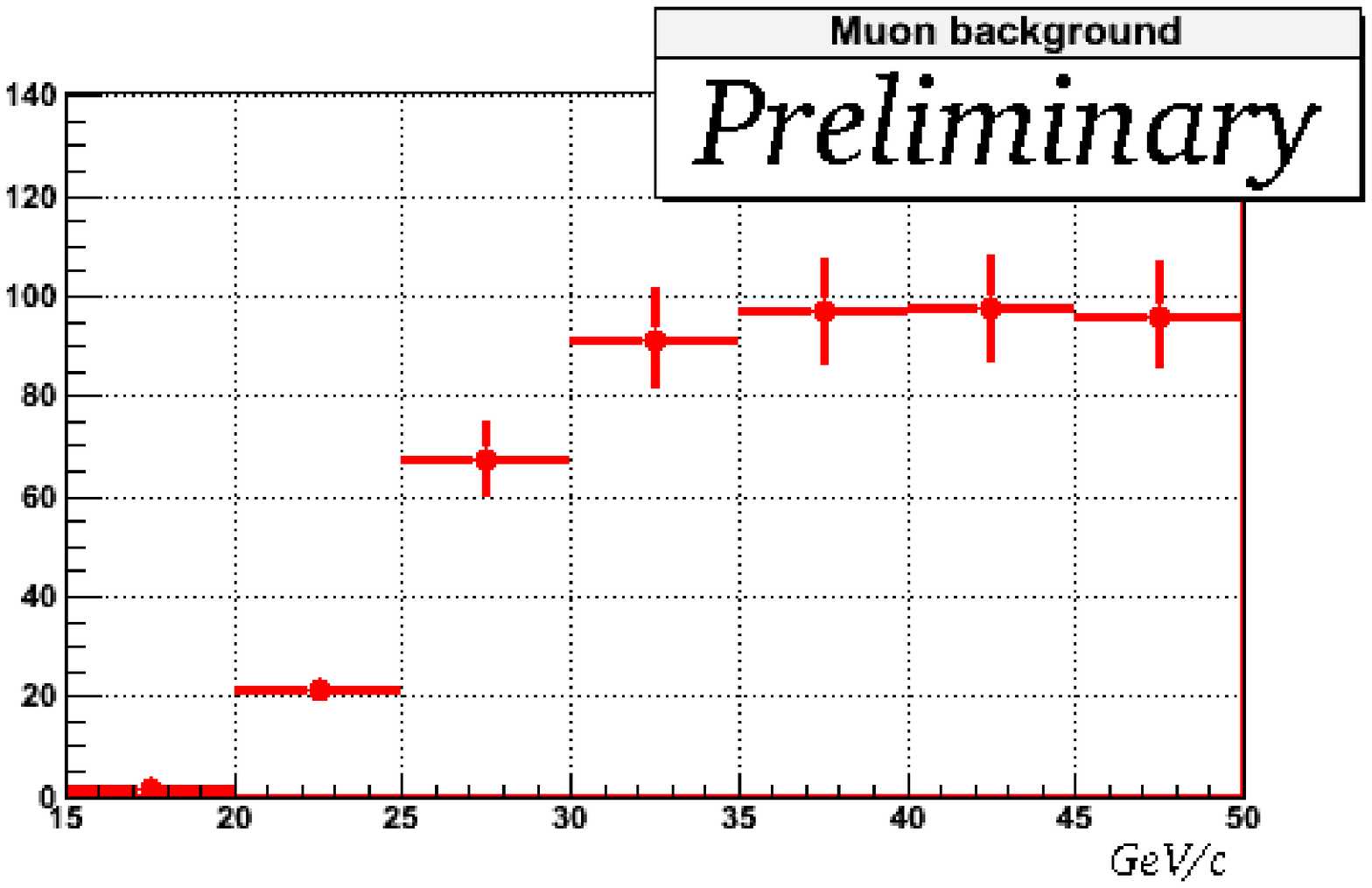}
    \caption{Number of \km background events in the \ke sample in each of the momentum bins}
    \label{ke2back}
  \end{minipage}\hfill
\end{figure}

The reconstructed $Ke2$ candidates are shown in Figure \ref{ke2mmiss}. With the presented selection about $3.4\times10^6$  \km and 3930 \ke candidates were found. 

\subsection{Background estimation}
The major contribution to the background in the \ke sample was identified to come from \km events with muons releasing almost all their energy in the LKR. 
This type of background was estimated from data. 
The requirement for $E/pc >0.95$ was changed to $E/pc<0.2$ assuring only muon contribution and the missing mass cut was kept. 
Then the obtained number of events was scaled with the probability that a muon gives $E/pc > 0.95$, which was also obtained from data. 
The resulting background in each of the momentum bins is shown in Figure \ref{ke2back}

After subtraction of all background sources the total number of \ke events was obtained to be $ 3407 \pm 63_{stat} \pm 54_{syst} $ 

The background in the \km sample was found to be approximately $0.6$\% dominated by $K^{\pm}\to \pi^{\pm}\pi^0$ events.

\subsection{Simulation}
A full GEANT3 \cite{geant3} based simulation of the detector response was used only for calculation of the acceptance for both decay modes. The radiative corrections were taken into account according to \cite{radiative} and \cite{Finkemeier}. 
The agreement between the data and Monte Carlo was very good as shown in Figure \ref{DATAMC}. 
%\begin{figure}[!htb]
%    \centering \includegraphics[width=90mm]{kmu2_data_mc.eps}
%	\caption{Data and MC distributions for the reconstructed vertex of $K\mu2$ events}
%	\label{DATAMC}
%\end{figure}

%\begin{figure}[!htb]
%  \begin{minipage}[t]{0.46\textwidth}
%    \centering \includegraphics[width=70mm]{kmu2_data_mc.eps}
%	\caption{Data and MC distributions for the reconstructed vertex of $K\mu2$ events}
%	\label{DATAMC}
%  \end{minipage}\hfill
%  \begin{minipage}[t]{0.46\textwidth}
%    \centering \includegraphics[width=70mm]{data_mc.eps}
%	\caption{Data and MC distributions for the reconstructed vertex of $K\mu2$ events}
%	\label{DATAMC}
%  \end{minipage}
%\end{figure}

\section{Results and conclusions}
The value of $R_K$ was calculated as 
\begin{equation}
R_{K} =\frac{N_{Ke2raw} - N_{Ke2back}}{TrEff(Ke2)~ Acc(Ke2) ~ C_{e}} \times \frac{Acc(K \mu 2) ~ C_{\mu }}{D~ (N_{K\mu 2raw} - N_{K\mu 2back})}
\end{equation}
where $Kl2raw$ are the $Kl2$ candidates passing the presented selection and $Kl2back$ is the background contribution, 
$C_l$ represents the correction the correction resulting from particle identification, 
$Acc(Kl2)$ is the geometrical acceptance from the MC simulation and 
$D=50$ is the downscaling for the \km events, $l=e,\mu$. 
All relevant values were obtained in momentum bins and, apart from the geometrical acceptance, from the data themselves. 

Most of the systematics cancel in the ratio of the partial widths. The residual uncertainties are presented in Table~\ref{tab:syst}. 
The major contribution comes from the determination of the background in the \ke sample which was obtained from data and error scales with the statistics. 

\begin{table}[!htb]
\begin{center}
\begin{tabular}{l|r}
\hline
Systematics source  & Uncertainty, $\Delta R_K/R_K$\\
\hline
Background subtraction in \ke &       1.59\%~~~~~~~~~\\
Electron identification  &       0.24\%~~~~~~~~~\\
\ke geometric acceptance (MC stat) &  0.3\%~~~~~~~~~\\
Trigger inefficiency &                  0.3\%~~~~~~~~~\\
\hline
\end{tabular}
\end{center}
\vspace{-5mm}
\caption{A summary of the major systematic uncertainties.}
\label{tab:syst}
\end{table}

\begin{figure}[!htb]
  \begin{minipage}[t]{0.46\textwidth}

    \centering \includegraphics[width=70mm]{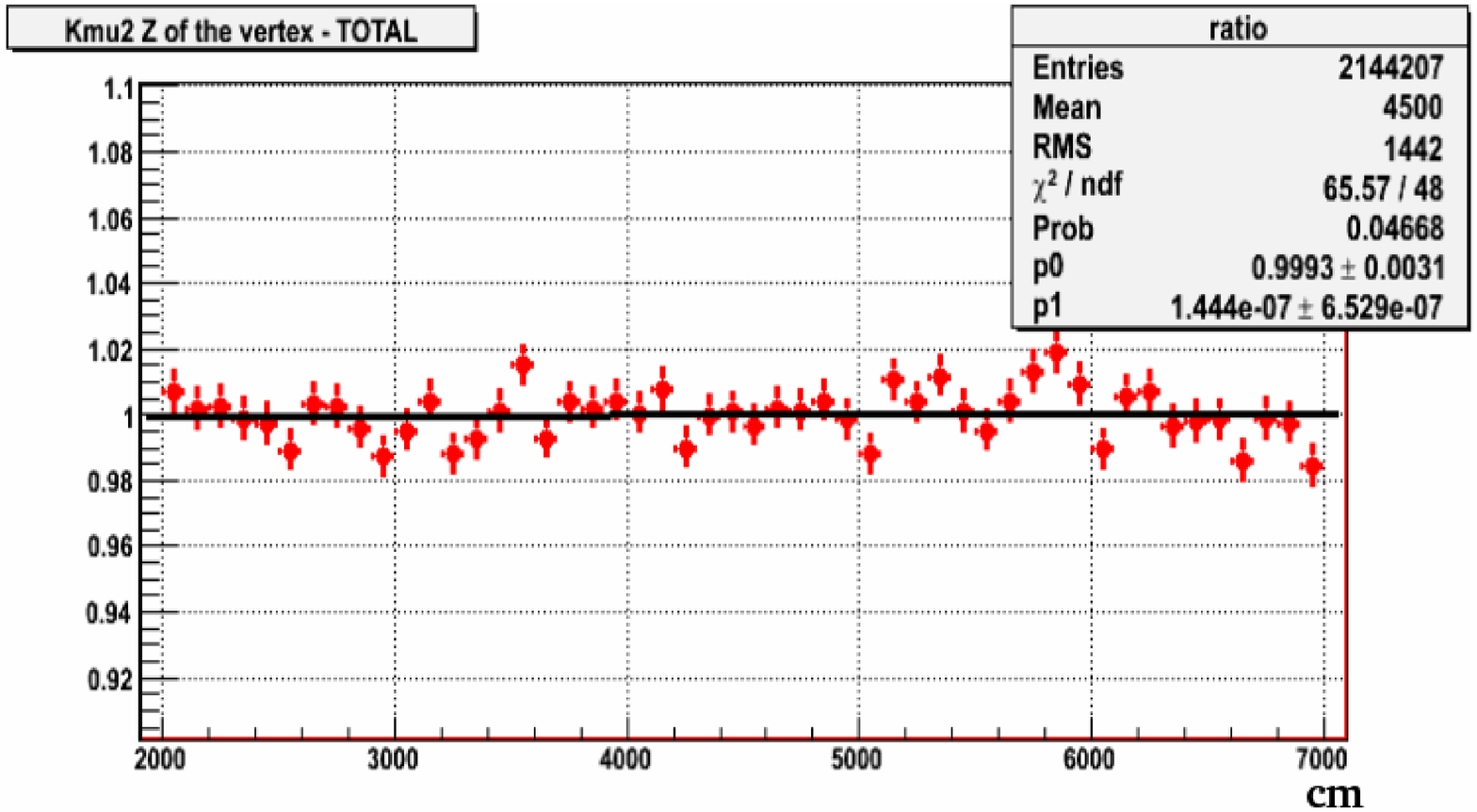}
	\caption{Ratio of the normalized data and MC distributions for the reconstructed Z-vertex coordinate of $K\mu2$ events}
	\label{DATAMC}

  \end{minipage}\hfill
  \begin{minipage}[t]{0.46\textwidth}

    \centering \includegraphics[width=70mm]{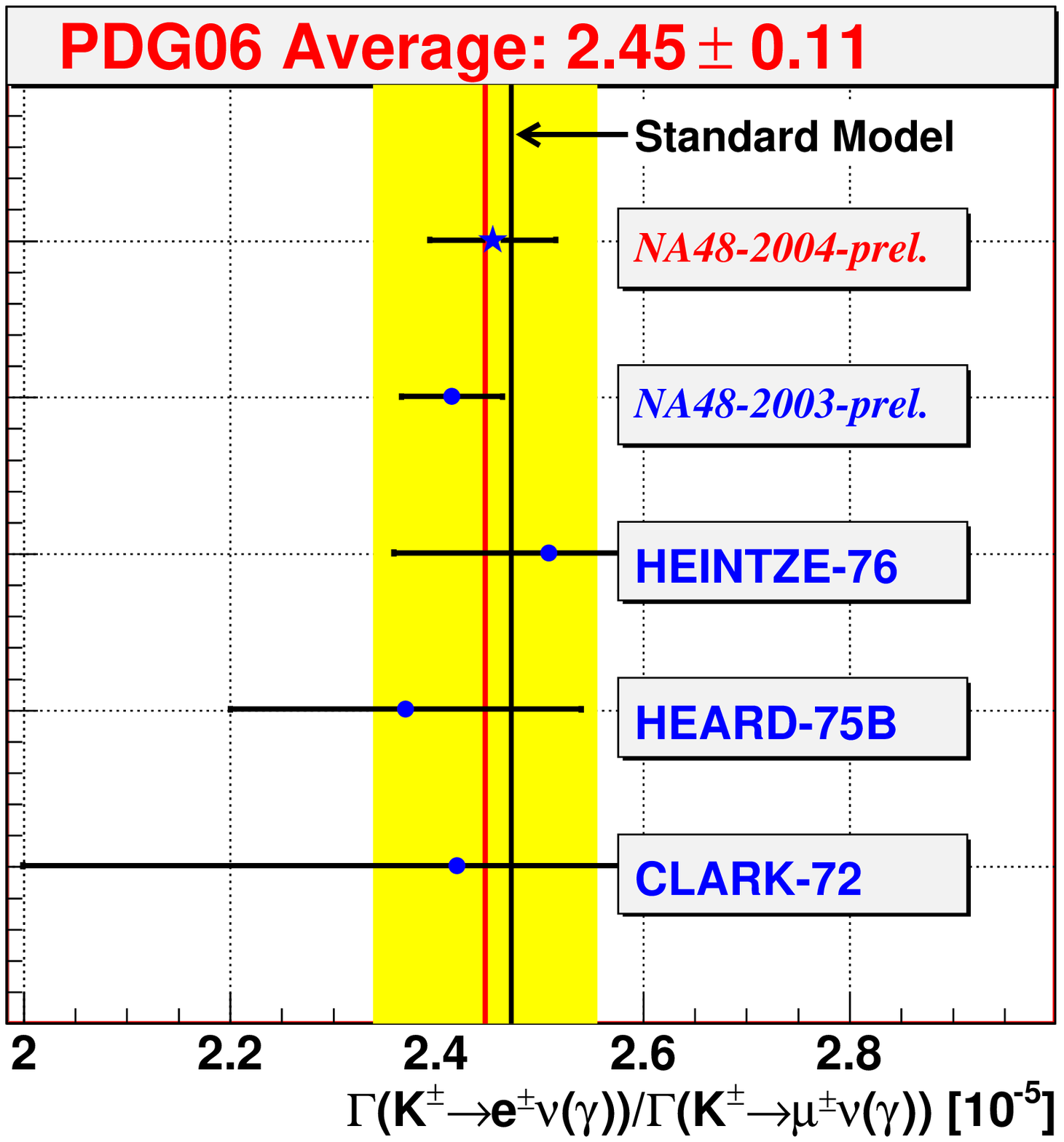}
	\caption{Different measurements of $R_K$. The average doesn't include the recent preliminary measurements by the NA48 experiment. The line representing the SM calculation includes the error}
	\label{rkall}
  \end{minipage}\hfill
\end{figure}

The preliminary result for $R_K$ from the 2004 data is 
\begin{equation}
R_{K} = (2.455 \pm 0.045 _{stat} \pm 0.041 _{syst}) \times  10^{-5}
\end{equation}
and the values for both kaon charges are consistent. 
The uncertainty is dominated by the statistics of the $Ke2$ sample and the total error is two times smaller 
than the combined result from the previous experiments as shown in Figure \ref{rkall}

The precision is still far from being competitive with the Standard Model prediction but is 
already very close to the level where non SM physics may give effects. 
Currently the NA48/2 collaboration is taking data to measure $R_K$ at the sub-percent level. 
Such measurement will provide a sensible test of the Standard Model and 
will allow to probe regions in the parameter space of some of its extensions.


\begin{thebibliography}{9}

\bibitem{Finkemeier} M. Finkemeier, Phys. Lett. B387 (1996), 391 
\bibitem{Masiero} A. Masiero et al., Phys. Rev. D74 (2006), 011701 
\bibitem{ke2prev} J. Heintze et al., Phys. Lett. 60 B (1976), 302; 
K.S. Heard et al., Phys. Lett. 55B (1975), 327; 
A.R. Clark et al.,
 Phys. Rev. Lett 
29 (1972), 1274 
\bibitem{na482} J.R.Batley et al., CERN/SPSC 2000-003 (2000) 
\bibitem{NA48setup}
  V.~Fanti {\it et al.}  [NA48 Collaboration],
  %``The Beam and detector for the NA48 neutral kaon CP violations experiment at
  %CERN,''
  Nucl.\ Instrum.\ Meth.\  A {\bf 574} (2007) 433.
  %%CITATION = NUIMA,A574,433;%%
\bibitem{geant3} 
CERN Program Library Long Write-up W5013 (1993)
\bibitem{radiative} J. Bijnens et al., Nucl. Phys B396 (1993), 81 

\end{thebibliography}
\end{document}